\documentclass[12pt]{article}

 \usepackage{amssymb}
 \usepackage{graphicx}  

\newcommand{\bigfrac}[2] {\frac{\textstyle #1}{\textstyle #2}}

\newcommand{\newc}{\newcommand}
\newc{\ra}{\rightarrow}
\newc{\lra}{\leftrightarrow}
\newc{\beq}{\begin{equation}}
\newc{\eeq}{\end{equation}}
\newc{\barr}{\begin{eqnarray}}
\newc{\earr}{\end{eqnarray}}

\def\al{\alpha}

\def\Braket#1#2#3{\langle{#1}|{#2}|{#3}\rangle}

\begin{document}
\begin{titlepage}
\begin{center}

{\large \bf  Nuclear matrix elements for exclusive
neutrino-nucleus reactions}

\vspace{12mm}

V.Ch. CHASIOTI AND T.S. KOSMAS
\vspace{2mm}

{\it Theoretical Physics Division, University of Ioannina,
GR-45110 Ioannina, Greece }
\end{center}
\vspace{7mm}

\centerline{\bf \today }

\vspace{7mm}

\begin{abstract}
The formalism describing (anti)neutrino-nucleus reaction
cross-sections in neutral- and charged-current processes is
improved. Compact formulae for the single-particle transition
matrix elements, based on the multipole expansion treatment of the
relevant hadronic currents and the use of harmonic oscillator
basis, are presented. As an application, the nucleus $^{127}\!I$,
a promising target for detection of solar- and supernova-neutrinos
as well as of cold dark matter candidates, is considered. Our
matrix elements refer to exclusive processes of the
charge-changing $^{127}\!I(\nu_l,l^-)^{127}\!Xe^{\star}$ reaction
and especially to the transition $\frac{5}{2}^+ \to \frac{3}{2}^+$
leading from the ground state of $^{127}$I to the lowest
excitation of $^{127}$Xe nucleus.
\end{abstract}

PACS number(s): 23.20.Js, 23.40.-s, 25.30.-c, 24.10.-i.
\vspace{0.5cm}

KEYWORDS:  Neutrino-nucleus interactions, lepton-induced
reactions, multipole matrix elements, harmonic oscillator basis,
quasiparticle-phonon model.

\end{titlepage}

\section{Introduction}

It is well known that, the neutrinos play a very significant role
in many phenomena in nature \cite{DonPe}-\cite{Jach99}. In spite
of their importance, however, numerous questions concerning their
properties \cite{Kol-Kos}, their roles in the solar-neutrino
puzzle \cite{Enge}, the atmospheric neutrino anomaly and the dark
matter problem \cite{KV97}, their oscillation-characteristics
\cite{Kos01,KRF00}, etc., remain still unanswered. During the last
decades, several ideas have been proposed to explain the
surprising behavior of neutrinos, but only a few of them have been
verified. Among others, this is due to the fact that the neutrinos
interact with matter only by means of the weak interactions. The
goal of the extensive experimental and theoretical investigations
related to neutrinos in nuclear physics, particle physics,
astrophysics and cosmology, is to shed light on the above open
issues.

Among the probes which involve neutrinos, the neutrino-nucleus
interaction possess a prominent position
\cite{Don-Wal}-\cite{Jach99}. Thus, the study of neutrino
scattering with nuclei is a good way to detect or distinguish
neutrinos of different flavor and explore the basic structure of
the weak interactions. Also, specific neutrino-induced transitions
between discrete nuclear states with good quantum numbers of spin,
isospin and parity allows us to study the structure of the weak
hadronic currents. Furthermore, terrestrial experiments performed
to detect astrophysical neutrinos (solar neutrinos, etc.) as well
as neutrino-induced nucleosynthesis interpreted through several
neutrino-nucleus interaction theories, constitute good sources of
explanation for neutrino properties \cite{Don-Wal}-\cite{Jach99}.

There are four categories of neutrino--nucleus processes: the two
types of charged-current (CC) reactions of neutrinos and
antineutrinos and the two types of neutral-current (NC) ones. In
the charged-current reactions a neutrino $\nu_l$ (antineutrino
$\bar {\nu}_l$) with  $l = e, \ \mu, \ \tau$, transforms one
neutron (proton) of a nucleus to a proton (neutron), and a charged
lepton $l^-$ (anti-lepton $l^+$) is emitted as
\begin{equation}
 \label{eq:cc}
\begin{array}{rcl}
  \nu_l + {}_{Z}A_{N} &\longrightarrow& {}_{Z+1}A^*_{N-1} + l^-
   \;, \\[1ex]
  \overline{\nu}_l + {}_Z A_N &\longrightarrow& {}_{Z-1}A^*_{N+1} + l^+
    \;.
 \end{array}
\end{equation}
These reactions are also called neutrino (anti-neutrino) capture,
since they can be considered as the reverse processes of
lepton-capture. They are mediated by exchange of heavy $W^\pm$
bosons according to the (lowest order) Feynman diagram shown in
Fig. 1(a). In neutral-current reactions (neutrino scattering) the
neutrinos (anti-neutrinos) interact via the exchange of neutral
$Z^0$ bosons [see Fig. 1(b)] with a nucleus as
\begin{equation}
 \label{eq:nc}
 \begin{array}{rcl}
  \nu + {}_{Z}A_{N} &\longrightarrow& {}_{Z}A^*_N + \nu^\prime
  \; , \\[1ex]
  \overline\nu + {}_{Z} A_N &\longrightarrow& {}_{Z}A^*_N +
  \overline\nu^\prime \;,
 \end{array}
\end{equation}
where $\nu$ ($\overline\nu$) denote neutrinos (anti-neutrinos) of
any flavor. The neutrino-nucleus reactions leave the final nucleus
mostly in an excited state lying below particle-emission
thresholds (semi-inclusive processes) \cite{Kos-Ose,Enge,Kolbe}.
The transitions to energy-levels higher than the particle-bound
states usually decay by particle emission and, thus, they supply
light particles that can cause further nuclear reactions
\cite{Kolbe}.

Theoretically, the various neutrino-nucleus transition rates are
calculated by following detailed prescriptions such those
described in Refs. \cite{Don-Wal}-\cite{Jach99} (for a review on
these methods see \cite{DonPe,Kol-Kos}). From the nuclear
structure viewpoint the main task is the evaluation of the matrix
elements of multipole operators between the initial and final
nuclear many-body states \cite{Solov}-\cite{KoSu}. In Ref.
\cite{HaKos}, by exploiting the multipole decomposition of the
hadronic currents, we have constructed a general closed formalism
which gives all types of the radial matrix elements required for
the description of any electromagnetic and weak process in nuclei.
This method uses harmonic oscillator single-particle basis and
allows the easy calculation of the basic radial matrix elements by
means of analytic formulas.

In the present work, we proceed a step further and extend the
method of Ref. \cite{HaKos} so as to provide similar expressions
for the reduced matrix elements of any basic single-particle
tensor operator entering the neutrino-nucleus cross-sections
\cite{Don-Wal}-\cite{Jach99}. As we shall see, the advantage of
this method lies in the separation of the geometrical coefficients
from the kinematical parameters of the reaction in question. In
this way, the transition matrix elements for every value of the
momentum transfer q can be evaluated in a direct way. We apply
this formalism in order to determine the needed matrix elements
for nuclei in the region of $^{127}\!I$ isotope, which is
considered a promising target for detection of solar and supernova
neutrinos \cite{Enge} as well as for detection of cold dark matter
candidates \cite{KV97}.

In the remainder of the paper, after a brief review of the
relevant expressions involved in the formal description of the
neutrino-nucleus cross sections and the multipole decomposition
treatment of the hadronic currents (Sect. 2), we proceed with the
derivation of closed analytic formulas for the basic
single-particle reduced matrix elements (Sect. 3). As an
application we calculate the transition matrix elements in the
case of the charged-current neutrino-nucleus reaction
$^{127}\!I(\nu_l,l^-)^{127}\!Xe^{\star}$ (Sect. 4). Finally, our
conclusions are summarized in Sect. 5.

\section{Weak interaction Hamiltonian of neutrino--nucleus reactions}

The effective Hamiltonian, describing the weak interaction of
nuclei with neutrinos at low energies compared to the electroweak
scale, can be written in current--current form as
\begin{eqnarray}
   \mathcal{ H}_{\rm eff} & = & \frac{G}{\sqrt2}
    \left(j_\lambda^{(-)}J^{\lambda(+)}+j_\lambda^{(0)}J^{\lambda(0)}
   +{\rm h.\ c.} \right) \; ,
   \label{eq:Heff}
\end{eqnarray}
($G$ denotes the weak interaction coupling constant) where
$j_\lambda^{(c)}$ and $J^{\lambda(c)}$ with $c = -,+,0$, denote
the leptonic and hadronic currents, respectively. For the
definition of the leptonic current see e.g. Ref. \cite{Kol-Kos}.
The expressions for the hadronic currents, since nucleons are
extended objects and Lorentz covariance should be fulfilled, reads
\cite{DonPe}
\begin{eqnarray}
   J_{\lambda}^{(c)} & = & \overline{\Psi}_N \left[
    g_1^V \ \gamma_{\lambda}
    + \frac{i}{2M} g_2^V \ \sigma_{\lambda \nu} q^{\nu}
    +              g_3^V \ q_{\lambda}   \right.
    \nonumber \\
  & + & \left.
    g_1^A \ \gamma_{\lambda} \gamma_5
    + \frac{i}{2M} g_2^A \ \sigma_{\lambda \nu} q^{\nu} \gamma_5
    +              g_3^A \ q_{\lambda} \gamma_5
     \right] \tau_c \Psi_N  \;,
\end{eqnarray}
($\Psi_N$ represent the nucleon isospin doublet) where the weak
nucleon form factors $g_i^V$, $g_i^A$ ($i=1,2,3$) are complex
scalar functions of the momentum transfer $q^2$. The $g_i^V $ are
fixed by the conserved vector-current theory (CVC) stating that
the isovector part of the electromagnetic current and the charge
raising and lowering parts of the weak vector current form an
isospin triplet of conserved currents \cite{DonPe}. For the axial
form factors $g_i^A $, charge symmetry properties and T-invariance
of the hadronic current require that $g_2^A  = 0$. Furthermore,
the $g_3^A q_{\lambda} \gamma_5$ term gives contributions
proportional to the mass of the outgoing lepton, and can therefore
be neglected in the extreme relativistic limit \cite{DonPe}. Thus,
one arrives at the expressions
\begin{eqnarray}
J^{(+)}_{\lambda} \,= \, \overline{\psi}_{N} \left\{ \frac{1}{2}
(F_1^p - F_1^n) \gamma_{\lambda} +  \frac{i}{4M} ( F_2^p - F_2^n )
\sigma_{\lambda \nu} q^{\nu} + G_A \gamma_{\lambda} \gamma_5
\right\} \tau_c \Psi_N \, , \label{Char-cur}
\end{eqnarray}
\begin{eqnarray}
J^{(0)}_{\lambda} & = & \overline{\psi}_{N} \left\{F_1^Z
\gamma_{\lambda} + F_2^Z \frac{i \sigma_{\lambda \nu} q^\nu}{2M} +
G_A \gamma_{\lambda} \gamma_5 \right\} \psi_{N} \;,
\label{Nutr-cur}
\end{eqnarray}
($J_\lambda^{(-)}$ is the Hermitian conjugate of
$J_\lambda^{(+)}$). Here $G_A = - \frac{1}{2}G_A^3 {\bf \tau_0}$
($\tau_0=+1$ for protons and $\tau_0=-1$ for neutrons) and
$F_{1,2}^p$, $F_{1,2}^n$ denote the charge and electromagnetic
form factors of proton and neutron (within the nucleus),
respectively. These form factors are discussed in Refs.
\cite{Kol-Kos,Kolbe}.

\subsection{The formalism for neutrino--nucleus cross sections}

By utilizing the neutrino--nucleus weak interaction Hamiltonian
discussed before, the neutrino scattering  cross sections can
reliably be calculated within the first-order Born approximation.
If we assume that the initial $\vert i\rangle$ and final $\vert
f\rangle$ nuclear states have well-defined spins and parities, a
multipole analysis of the weak (nuclear-level) hadronic current
can be performed \cite{HaKos} (see below Sect. 2.2). This has been
carried out in close analogy to electron scattering from nuclei
\cite{CNP,KV92} within a unified analysis of charge-changing
semi-leptonic weak interactions in nuclei.

The differential (with respect to energy and scattering direction)
neutrino--nucleus scattering cross-section is written as
\cite{Kol-Kos}
\begin{eqnarray}
\label{eq:Sec2_1}
   \frac{{\rm d}^{2} \sigma_{i \rightarrow f}}
    {{\rm d} \Omega {\rm d} \omega } & =
& \frac{G^2}{\pi} \frac{|{\bf k}_f| \ \epsilon_f}{(2J_i+1)} \,
  F(Z,\epsilon_f) \left(\sum \limits_{J=0}^\infty \sigma_{CL}^J +
  \sum \limits_{J=1}^\infty \sigma_{T}^J \right) \;,
\end{eqnarray}
where $\omega =\epsilon_i- \epsilon_f$ is the excitation energy of
the nucleus, and $\epsilon_i$, $\epsilon_f$, ${\bf k}_f$ denote
the energy of the incoming neutrino and energy and momentum of the
outgoing lepton, respectively. The summations in Eq.
(\ref{eq:Sec2_1}) contain the contributions of the Coulomb
($\widehat{\mathcal{ M}}_J$), longitudinal ($\widehat{\mathcal{
L}}_J$), transverse electric ($\widehat{\mathcal{T}}_J^{el }$) and
transverse magnetic ($\widehat{\mathcal{T}}_J^{mag}$) operators
stemming from the multipole expansion of the weak hadronic
current. The operators $\widehat{\mathcal{ M}}_J$,
$\widehat{\mathcal{ L}}_J$, $\widehat{\mathcal{T}}_J^{el}$ and
$\widehat{\mathcal{T}}_J^{mag}$ contain both polar-vector and
axial-vector parts of which the contributions are written as
\begin{eqnarray}
\label{eq:Sec2_2}
   \sigma_{CL}^J & = & \left( 1 + a \cos{\Phi} \right)
   \left| \Braket{J_f}{|\widehat{\mathcal{ M}}_J(q)|}{J_i} \right|^2
+ \left( 1 + a \cos{\Phi} - 2 b \sin^2{\Phi} \right)
   \left| \Braket{J_f}{|\widehat{\mathcal{ L}}_J(q)|}{J_i} \right|^2
\nonumber \\[1ex]
&+&  \left[ \bigfrac{\omega}{q} \left( 1 + a \cos{\Phi} \right) + d \right]
     2 \Re e \Braket{J_f}{|\widehat{\mathcal{ L}}_J(q)|}{J_i}
     \Braket{J_f}{|\widehat{\mathcal{ M}}_J(q)|}{J_i}^\ast \;,
    \\[2ex]
\label{eq:Sec2_3}
   \sigma_{T}^J & = & \! \!
   \left( 1 \! - \! a \cos{\Phi} + b \sin^2{\Phi} \right)
\left[\left| \Braket{J_f}{|\widehat{\mathcal{T}}_J^{mag}(q)|}{J_i}
\right|^2
 + \left| \Braket{J_f}{|\widehat{\mathcal{T}}_J^{el}(q)|}{J_i} \right|^2
                       \right] \nonumber \\[1ex]
& \mp & \! \! \left[ \frac{(\epsilon_i \! + \! \epsilon_f)}{q} \!
     \left( 1 \! - \! a \cos{\Phi} \right) \! - \! d \right] \!
2 \Re e \! \Braket{J_f}{|\widehat{\mathcal{T}}_J^{mag}(q)|}{J_i}
\! \Braket{J_f}{|\widehat{\mathcal{T}}_J^{el}(q)|}{J_i}^\ast \;,
\end{eqnarray}
where $\Phi$ denotes the scattering angle of the outgoing lepton and
$a$, $b$ and $d$ are given by
\begin{equation}
\label{eq:Sec2_4} a = \frac{|{\bf k}_f|}{\epsilon_f}
  = \left[ 1-\left( \frac{m_f c^2}{\epsilon_f} \right)^2 \right]^
    {\frac{1}{2}} \;, \quad
b = \frac{\epsilon_i \epsilon_f a^2}{q^2} \;, \quad
d = \frac{\left( m_f c^2 \right)^2}{q \  \epsilon_f} \;,
\end{equation}
($m_f$ is the mass of the outgoing lepton). The magnitude of the
three-momentum transfer $q$ is given by
\begin{eqnarray}
q &=& |\!{\,{\bf q}}|  =  \left[\omega^2 +
   2 \epsilon_i \epsilon_f \left( 1 - a \cos{\Phi} \right)
   - \left( m_f c^2 \right)^2\right]^{\frac{1}{2}} \;.
   \label{eq:Sec2_5}
\end{eqnarray}
Notice that the interference term between vector and axial vector
current in Eq. (\ref{eq:Sec2_3}) has a negative (positive) sign
for neutrino (antineutrino) scattering due to their different
helicities.

The well-known Fermi function $F(Z,\epsilon_f)$ takes into account
the Coulomb--final-state interaction between nucleus and final
lepton (case of charged-current reactions only).

\subsection{ Multipole--decomposition operators }

The standard multipole expansion procedure \cite{DonPe}, applied
on the matrix elements of hadronic current $\hat J_{\mu}({\bf r})
= (\hat {\rho}, \hat {\bf J})$, where $\hat {\rho}({\bf r})$ the
density and $\hat {\bf J}({\bf r})$ the three-current operators,
leads to spherical tensor operators which are given in terms of
the projection functions
\beq
M^J_M(q{\bf r}) \,=\,\delta_{LJ} j_L(qr)Y^L_M(\hat r),
\label{proj1M-M}
\eeq
\beq  {\bf M}_M^{(L1)J}(q{\bf r}) \, =\, j_L(qr){{\bf
Y}}_M^{(L1)J}(\hat r) \label{proj2M-M} \, . \eeq
Here $j_L(r)$ stands for the spherical Bessel functions,
$Y^L_M({\hat r})$ are the spherical Harmonics and ${\bf
Y}^{(L,1)J}_M({\hat r})$ are the vector spherical Harmonics. In
the case of the polar-vector current ${\hat J}_\lambda$, the
multipole decomposition gives the operators $\hat M_{JM}^{coul}$,
$\hat L_{JM}$, $\hat T_{JM}^{el}$ and $\hat T_{JM}^{mag}$. The
first three operators have parity $(-)^J$ (normal parity
operators), while the parity of $\hat T_J^{mag}$ is $(-)^{J+1}$
(abnormal parity operator). In the case of axial-vector current
${\hat J}_{\lambda}^5$ we obtain the operators $\hat M_{JM}^{Coul
5}$, $\hat L_{JM}^5$, $\hat T_{JM}^{el5}$ and $\hat
T_{JM}^{mag5}$. The first three axial-vector multipoles have
parity $(-)^{J+1}$ while $\hat T_J^{mag5}$ is a normal parity
operator.

For a conserved vector current (CVC) like the electromagnetic, $
\hat L_{JM_J}(q)= (q_0/q) {\hat M}^{Coul}_{JM_J}(q)$, where $q_0$
represent the time component of the four-momentum transfer, $q_\mu
= (q_0, {\bf q}$). In this case, the number of independent
operators resulting from the decomposition procedure is reduced to
seven. In fact, the matrix elements of these seven basic operators
involve isospin dependent form factors $F_X^{(T)}(q^2)$ (see Eqs.
(\ref{Char-cur}), (\ref{Nutr-cur}) and Ref. \cite{HaKos}). For
this reason, we define seven new operators as \cite{DonPe}
\barr T_1^{JM} \, \equiv \, M^J_M(q{\bf r}) \, = \, \delta_{LJ}\,
j_L(\rho)Y^L_M(\hat r),
\label{opM}
\earr
\barr T_2^{JM} \,
\equiv\, {\Sigma}^J_{M}(q{\bf r}) \,=\,{{\bf M}}^{JJ}_{M}\cdot
 \mbox{\boldmath $ \sigma $},
\label{opSig}
\earr
\barr
T_3^{JM} \, \equiv \, {\Sigma^{\prime}}^J_{M}(q{\bf r})  \,= \,
-i\left\{\frac{1}{q}
\nabla \times {{\bf M}}^{JJ}_{M}(q{\bf r})\right\} \cdot
 \mbox{\boldmath $ \sigma $},
\label{opSigp}
\earr
\barr
T_4^{JM} \, \equiv \,{\Sigma  ''}^J_{M}(q{\bf r}) \, = \,
 \Big\{\frac{1}{q}\nabla
M^J_{M}(q{\bf r})\Big\} \cdot
\mbox{\boldmath  $ \sigma $},
\label{opSigpp}
\earr
\barr
T_5^{JM} \, \equiv \,{\Delta}^J_{M}(q{\bf r}) \,= \,
{{\bf M}}^{JJ}_{M}(q{\bf r})
\cdot\frac{1}{q}\nabla,
\label{opDel}
\earr
\barr
T_6^{JM} \, \equiv \,{\Delta ^{ \prime}}^J_{M}(q{\bf r}) \,= \,
-i\Big\{\frac{1}{q}\nabla \times {{\bf M}}^{JJ}_{M}(q{\bf r})\Big\}
 \cdot \nabla,
\label{opDelp} \earr \barr T_7^{JM} \, \equiv \,
\Omega^J_{M}(q{\bf r}) \, = \, M^J_{M}(q{\bf r})\mbox{\boldmath  $
\sigma $} \cdot \frac{1}{q}\nabla. \label{opOm}
\earr

Using properties of the nabla operator ($\nabla$), Eqs.
(\ref{opSigp}), (\ref{opSigpp}) and (\ref{opDelp}) can be
rewritten as
\beq T_3^{JM} \,
\equiv\,{\Sigma^{\prime}}^J_{M} \, = \, [J]^{-1} \Big\{-J^{1/2}
{\bf M}^{J+1J}_{M} + (J+1)^{1/2} {\bf M}^{J-1J}_{M}\Big\} \cdot
 \mbox{\boldmath $ \sigma $},
\label{opSigpt} \eeq \beq T_4^{JM}
\,\equiv\,{\Sigma^{\prime\prime}}^J_M \, = \,[J]^{-1}
\Big\{(J+1)^{1/2}{{\bf M}}^{J+1J}_{M} + J^{1/2}{{\bf
M}}_{J-1J}^{M}\Big\} \cdot \mbox{\boldmath  $ \sigma $},
\label{opSigppt} \eeq \beq T_6^{JM} \, \equiv\,{\Delta ^{
\prime}}^J_{M} \,= \,[J]^{-1} \Big\{-J^{1/2}{{\bf M}}^{J+1J}_{M} +
(J+1)^{1/2}{{\bf M}}^{J-1J}_{M}\Big\} \cdot \frac{1}{q}\nabla,
\label{opDelpt} \eeq (throughout this work we use the common
symbol $[J] = (2J+1)^{1/2}$).

Many quantities describing the semi-leptonic electroweak processes
in nuclei, including, of course, the neutrino-nucleus reactions,
are expressed (to a good approximation) in terms of
single-particle nuclear matrix elements of the one-body operators
$T_i^{JM}$, i=1,2,...7 \cite{DonPe}. By applying the Wigner-Eckart
theorem, these matrix elements are written in terms of the
following four reduced matrix elements:
$\Braket{j_1}{|M^J|}{j_2}$, $\Braket{j_1}{|{{\bf
M}^{LJ}\cdot\mbox{\boldmath $\sigma$}}|}{j_2}$,
$\Braket{j_1}{|{\bf M}^{LJ}\cdot (\nabla /q)|}{j_2}$ and
$\Braket{j_1}{|M^L\mbox{\boldmath$\sigma$}\cdot (\nabla
/q)|}{j_2}$. In the next section, we will use harmonic oscillator
basis in order to find simplified expressions for these reduced
matrix elements.

\section{The Single-particle reduced matrix elements }

By applying the re-coupling relations and using the formalism of
Ref. \cite{HaKos}, the reduced matrix elements of the operators
${\mathcal T}_i^J$, $i=1,2,...7$, can be written in the compact
forms shown below.

1. For the operators $M^J_M \equiv {\mathcal O}_1^{JM}$ and ${\bf
M}^{LJ}_M\cdot\mbox{\boldmath $\sigma$} \equiv {\mathcal
O}_2^{JM}$ the reduced matrix elements $\Braket{j_1}{|{\mathcal
O}_i^{J}|}{j_2}$, have previously been written as
\cite{KV97,KRF00}
\beq
\Braket{j_1}{|{\mathcal O}_i^{(L,S_i)J}|}{j_2}\,= \,
(l_1 \  L \  l_2)\,{\mathcal U}^J_{LS_i}
 \Braket{n_1l_1}{j_L(\rho)}{n_2l_2}, \quad i=1,2 \, ,
\label{gen-ME}
\eeq
(with $S_1=0$ and $S_2=1$) where the symbols $(l_1 \ L \ l_2)$ and
${\mathcal U}_{LS}^J$ contain the 3-j and 9-j symbols,
respectively (see Ref. \cite{HaKos}).

2. The reduced matrix elements of $\Braket{j_1}{|{\bf
M}^{LJ}(q{\bf r}) \cdot (\nabla /q)|}{j_2}$, after some
manipulation can be cast in the form
\beq \Braket{j_1}{|{\bf M}^{LJ}(q{\bf r}) \cdot \frac{1}{q}
\nabla|}{j_2} \,=\, \sum_{\al} {\mathcal A}_L^{\al}(j_1 j_2 ;J)
\Braket{n_1l_1}{\theta _L^{\al}(\rho)}{n_2l_2}, \quad \al = \pm \,
, \label{MdelME} \eeq
where
\barr {\mathcal A}_L^{\pm}(j_1 j_2; J) \, & = &
\,\pm(-)^{l_1+L+j_2+1/2} \left(\frac{2l_2+1 \mp 1}{2}\right)^{1/2}
[j_1][j_2][J] (l_1 \  L \  l_2 \mp 1)
\nonumber \\
&& \times {\mathcal W}_6(l_1, j_1, 1/2, j_2, l_2, J) {\mathcal
W}_6( L, 1, J, l_2, l_1, l_2 \mp 1) \, . \label{CfMdelME} \earr
Here ${\mathcal W}_6$ represent the common 6-j symbol.

3. Similarly, for the  reduced matrix element of
$\Braket{j_1}{|M^J(q{\bf r})\mbox{\boldmath $\sigma$}
 \cdot (\nabla /q)|}{j_2} $ we can write
\beq
\Braket{j_1}{|M^J(q{\bf r})\mbox{\boldmath $\sigma$}
 \cdot \frac{1}{q} \nabla|}{j_2} \,= \,
\sum_{\al} {\mathcal B}_L^{\al}(j_1 j_2; J) \Braket{n_1l_1}{\theta
_J^{\al}(\rho)}{n_2l_2}, \quad \al = \pm \, ,\label{MspdelME} \eeq
where \barr {\mathcal B}_L^{\pm}(j_1 j_2; J) \,& = &\,\pm
\,{\delta}_{j_2,l_2\mp{1/2}}\, [j_1][j_2] (l_1 \  J \  2j_2 -l_2)
\, {\mathcal W}_6(l_1, j_1, 1/2, j_2, 2j_2-l_2, J) \, .
\label{CfMspdelME} \earr

From Eqs. (\ref{gen-ME}), (\ref{MdelME}) and (\ref{MspdelME}) we
notice that all basic single-particle reduced matrix elements
required for our purposes rely on the following three types of
radial integrals:
\beq \Braket{n_1l_1}{\theta^{\al}_l(\rho)}{n_2l_2} \,\equiv \,\int
dr r^2 R_{n_1l_1}^{*}(r)\theta^{\al}_l(\rho)R_{n_2l_2}(r),
\hspace{1.5cm} \al \, = \,0,\pm \, , \label{rad-int} \eeq
with \beq \theta^0_l(\rho) = j_l(\rho) \label{theta0} \, , \eeq
\beq \theta^{\pm}_l(\rho)  = j_l(\rho)\left(\frac{d}{d\rho} \pm
\frac{2l_2+1 \pm 1}{2\rho}\right). \label{theta} \eeq
The argument $\rho$ in Eqs. (\ref{theta}) is equal to $\rho = qr$.

By inserting in Eqs. (\ref{gen-ME}), (\ref{MdelME}) and
(\ref{MspdelME}), the expressions  for the radial matrix elements
found in Ref. \cite{HaKos} (see Appendix A) and manipulating
properly the appearing summations, we can write the four types of
reduced matrix elements entering the basic operators
(\ref{opM})-(\ref{opOm}) in closed forms, as follows:

1. For the Fermi- (S=0) and Gamow--Teller-type (S=1) operators
${\mathcal O}_i^{(L,S_i)J}$, we have
\beq \Braket{j_1}{|{\mathcal O}_i^{(L,S_i)J}|}{j_2}\, = \,
e^{-y}y^{L/2} \, \sum^{n_{max}}_{{\mu}=0} \,E_{\mu}^i(L) \,
y^{\mu} \, , \label{ME-Fe-GaT} \eeq
\beq
 y\,=\,(qb/2)^2 \, ,
\label{psi} \eeq
with $S_1=0$ and $S_2=1$, and where \beq E_{\mu}^i(L) \,=\, ( l_1
\,L \,l_2)\, {\mathcal U}_{LS_i}^J \,
{\varepsilon}_{\mu}^L(n_1l_1n_2l_2) \, \quad i=1,2 \, .
\label{cf-Eps} \eeq
2. The reduced matrix element $\Braket{j_1}{|{\bf M}^{LJ}(q{\bf
r}) \cdot (\nabla /q)|}{j_2}$ take the form
\beq \Braket{j_1}{|{\bf M}^{LJ}(q{\bf r}) \cdot
\frac{1}{q} \nabla|}{j_2} \, = \,
e^{-y}y^{(L-1)/2}\sum_{\mu=0}^{n_{max}}E_{\mu}^3(L) \, y^{\mu},
\label{ME-Mdel} \eeq
with
\beq E_{\mu}^3(L)\,=\, {\mathcal
A}_L^-\, \zeta_{\mu}^-(L) + {\mathcal A}_L^+ \,\zeta_{\mu}^+(L).
\label{cf-Zet} \eeq
3. Similarly for $\Braket{j_1}{|M^L(q{\bf r})\mbox{\boldmath
$\sigma$} \cdot \frac{1}{q} \nabla|}{j_2}$ we derive the formula
\beq \Braket{j_1}{|M^L(q{\bf r})\mbox{\boldmath $\sigma$} \cdot
\frac{1}{q} \nabla|}{j_2} \,  =  \,
e^{-y}y^{(L-1)/2}\sum_{\mu=0}^{n_{max}}E_{\mu}^4(L) \, y^{\mu},
\label{ME-Mspdel} \eeq
where
\beq E_{\mu}^4(L) \, = \,{\mathcal B}_L^-\, \zeta_{\mu}^-(L)
+ {\mathcal B}_L^+\, \zeta_{\mu}^+(L). \label{cf-Phi} \eeq
The coefficients $\varepsilon_{\mu}^L$ and $\zeta_{\mu}^{\pm}(L)$
are defined in Ref. \cite{HaKos} (see also Appendix A).

Having available the formalism of Eqs. (\ref{ME-Fe-GaT}),
(\ref{ME-Mdel}) and (\ref{ME-Mspdel}), we can straightforwardly
deduce closed analytic formulas for the reduced matrix elements of
the seven basic operators Eqs. (\ref{opM})-(\ref{opOm}). All these
fundamental matrix elements can be cast in the compact form
\beq \Braket{j_1}{|T^J|}{j_2}\, = \, e^{-y}y^{\beta/2}
\sum^{n_{max}}_{{\mu}=0} \,{\mathcal P}_{\mu}^J \, y^{\mu} \, ,
\label{gen-pol} \eeq
where \beq
 n_{max}=(N_1+N_2-\beta)/2 \, .
\label{nmax}
\eeq
The coefficients ${\mathcal P}_{\mu}^J$ for each specific case are
listed in Table 1. The integer $\beta$ of Eqs. (\ref{gen-pol}) and
(\ref{nmax}) is also quoted in this Table. As can be seen from
Table 1, the coefficients ${\mathcal P}_{\mu}^J$ are simply
related to the numbers $E_{\mu}^i(L)$, $i=1,2,3,4$, given in Eqs.
(\ref{cf-Eps}), (\ref{cf-Zet}) and (\ref{cf-Phi}). The
coefficients ${\mathcal P}^{\mu}_0$ are much simpler compared to
those corresponding to $\mu >0$.

We should remark that, upon mutual interchanging $n_1(l_11/2)j_1$
with $n_2(l_21/2)j_2$ in Eq. (\ref{gen-pol}), the following
relation holds: \beq \Braket{j_2}{|T^J(q{\bf r})|}{j_1} \,=\,
(-)^{\lambda}\Braket{j_1}{|T^J(q{\bf r})|}{j_2}, \label{prop-RME}
\eeq
where $\lambda \, = \,j_1-j_2$ for the operators M, $\Delta$,
$\Sigma^{\prime}$,  $\Sigma^{\prime \prime}$, and $\lambda \, =
\,j_1+j_2$ for the operators $\Delta^{\prime}$ and $\Sigma$ (our
phase convention for the matrix elements in Eq. (\ref{prop-RME})
is the same with that of Ref. \cite{KRF00}). The operator $\Omega$
does not have a simple phase symmetry under the above interchange,
$j_1 \leftrightarrow j_2$, and, for this reason, one traditionally
defines the operator \cite{DonPe}
\beq {\Omega^{\prime}}^J_M \,\equiv \, \Omega^J_M
+\frac{1}{2}{\Sigma^{\prime \prime}}^J_M \, , \label{opOmp} \eeq
for which, on applying Eq. (\ref{prop-RME}), one must put $\lambda
=j_1 +j_2$. The matrix elements of the operator ${\Omega}^{\prime
J}$ are also given by Eq. (\ref{gen-pol}) but now the coefficients
${\mathcal P}_{\mu}^J$ are (see Table 1)
\beq {\mathcal P}^J_{\mu} \,=\, E_{\mu}^4(J) +\frac{1}{2}
\big\{J^{1/2}E_{\mu}^2(J-1) + (J+1)^{1/2}E_{\mu -1}^2(J+1)\big\}.
\label{cf-Omp} \eeq

Before presenting some applications on the above formalism, it
should be noted that, the explicit and general expressions of Eq.
(\ref{gen-pol}) hold for every combination of the single-particle
levels $(n_1l_1)j_1$, $(n_2l_2)j_2$. For the determination of the
involved polynomials it suffices the evaluation of the geometrical
momentum independent coefficients ${\mathcal P}^J_{\mu}$ (see
Table 1). These closed formulas have been derived by inverting
properly the multiple summations involved in the corresponding
matrix elements (see Ref. \cite{HaKos,HaKo02a}), so as the final
summation is performed over the major harmonic oscillator quantum
number $N=2n+l$. It is worth remarking that such a compact
formalism provides a very useful insight to those authors who wish
to adopt a phenomenological approach and fit the nuclear
transition strengths for many semi-leptonic processes in nuclei
\cite{HaKos,HaKo02a}.

\section{Discussion, analysis and results}

Our primary aim in the present paper, is to deal with the
improvement of the formalism for the neutrino-nucleus reaction
cross sections. We shall focus our discussion on those reactions
that pertain to the terrestrial observation of astrophysical
neutrinos and especially on exclusive transitions leading to a
definite final state in neutrino-induced processes which have
attracted considerable interest in the past few years
\cite{Kol-Kos,Enge,Vol00}. A special category of such processes is
the one occurring in the radiochemical experiments leading to
excited states below particle emission thresholds, i.e. to the
particle-bound states (semi-inclusive processes). As an example of
this category, the reaction $\nu_l +^{127}\!I\ra^{127}\!Xe
^{\star} +l^-$ is discussed in some detail below.

As it is known, solar or supernova neutrinos, incident on an iodine
liquid target, could produce the noble gas $^{127}\!Xe$ which can
be recovered and counted as in the $^{37}\!Cl$ (radiochemical)
experiment. This suggests that, the reaction
\begin{equation}
\nu_l +^{127}\!I\ra^{127}\!Xe^{\star}+l^-\label{I-Xe} \,
\end{equation}
may be a promising radiochemical-type detector for $^7\!Be$ or
$^8\!B$ neutrinos \cite{Kos-Ose,Enge}. In the above reaction, the
ground state to ground state ($g.s.\to g.s.$) transition
$\frac{5}{2}^+ \ra\frac{1}{2}^-$, is forbidden, so that, the first
allowed one is the $\frac{5}{2}^+\ra \frac{3}{2}^+$ leading to the
lowest excitation (at 124.6~KeV) of $^{127}\!Xe$ isotope.

On the theoretical nuclear physics side, the many-body nuclear
wave-functions for heavy odd-A nuclei with relatively high
half-integer (ground state) spins, cannot be easily constructed
within the context of shell model or random phase approximation
methods. The medium heavy nuclei $^{127}\!I$ and $^{127}\!Xe$
which enter the reaction (\ref{I-Xe}) are isotopes of this sort.
For this reason, the wave-functions involved in the cross sections
of process (\ref{I-Xe}) are currently calculated \cite{KoSu}
within the framework of the microscopic quasiparticle-phonon
model. This model, has recently been used for reliable
descriptions of other similar processes \cite{Solov,Hul-Su}.

In the quasi-particle phonon (QPM) model \cite{Solov}, the wave
functions of odd-mass spherical nuclei are expressed in the form
of expansions in a phonon-basis. Thus, the main idea of the model
is to construct this phonon-basis, whereby phonons are defined as
solutions of the quasi-particle random phase approximation
\cite{Kos01}. The goal of the model is to be able to describe the
interplay between collective and single-particle degrees of
freedom of low-lying excited states. The effective Hamiltonian in
the QPM is written as
\begin{equation}
H_{eff} = H_{sp} + H_{pair} + H_{m} + H_{sm} \, , \label{H-QPM}
\end{equation}
where $H_{sp}$ is the single-particle Hamiltonian, $H_{pair}$
represents the monopole pairing interaction in the particle-hole
({\it p-h}) channel, $H_{m}$ is the separable multipole interaction in
the {\it p-h} channel, and $H_{sm}$ is the separable spin-multipole
interaction in the {\it p-h} channel.

To calculate the spectrum of the needed excited states for the
nuclei $^{127}\!I$ and $^{127}\!Xe$, the effective Hamiltonian Eq.
(\ref{H-QPM}) must be transformed in such a way that, instead of
the nucleon degrees of freedom, other simple modes of excitation
would be present. These simple modes are Bogolubov's
quasiparticles and phonons (that is why the model has been called
quasiparticle-phonon model). The required procedure for these
transformations is described in detail in Ref. \cite{Solov,KoSu}.

In the case of the neutral-current neutrino-$^{127}\!I$
scattering, one needs either the $g.s \to g.s.$ transition
(coherent process) of $^{127}\!I$, i.e. $\frac{5}{2}^+\ra
\frac{5}{2}^+$, or the transitions going from the ground state of
$^{127}\!I$ to its low-lying (particle-bound) states. In general,
under the assumptions discussed in the Introduction, any partial
cross section is written in terms of the square of the matrix
elements of tensor multipole operators (see Sect. 2) of the form
\begin{equation}
\Braket{f}{|T^J(q{\bf r})|}{i} \,=\, \sum_{j_1,j_2}
\Braket{j_1}{|T^J(q{\bf r})|}{j_2} \, D (j_1,j_2; J)
\label{Cr-Sec} \, .
\end{equation}
where $j_1,j_2$ run over the configurations of the chosen model
space which couple to a given J. The functions $D(j_1,j_2; J)$ are
the one-body transition densities which in our case are provided
by the quasiparticle-phonon model. The chosen model space in
\cite{KoSu} consists of the oscillator major shells with $N=3-5$
plus the intruder $0i_{13/2}$ from the $N=6$ major shell.

The single-particle reduced matrix elements
$\Braket{j_1}{|T^J(q{\bf r})|}{j_2}$ are evaluated by the
expression (\ref{gen-pol}). In Tables 2-4 we quote some
coefficients ${\mathcal P}_{\mu}^J$ needed to determine the
reduced matrix elements for the normal-parity operators $M^J,
\Sigma^J, \Delta ^{\prime J}$ (see Sec. 2.2) by using Table 1 and
Eq. (\ref{gen-pol}). Due to space limitations, in these Tables we
list the coefficients for configuration only of the type
$\frac{5}{2}^+ \ra \frac{3}{2}^+$ (for others the reader is
referred to Refs. \cite{HaKos,HaKo02a}).
As can be seen from Tables 2-4, the coefficients of the
polynomials involved in Eq. (\ref{Cr-Sec}) are very simple and
readily calculable numbers a fact that illustrates the advantage
of the formalism presented in Sect. 2-3.

At this point it is worth remarking that, regarding the neutrino
energies $E_{\nu}$ one is interested in, two main treatments can
be followed: (i) Include in Eq. (\ref{Cr-Sec}) the momentum
dependence of the multipole operators Eqs.
(\ref{opM})-(\ref{opOm}), and (ii) Reduce the multipole operators
to their $q\ra 0$ limit (long wave-length approximation)
\cite{DonPe}. It should be mentioned that, according to their
energies the neutrinos are classified in low-energy, $E_{\nu}\leq
20 MeV$ (solar neutrinos, low-energy supernova neutrinos),
medium-energy, $20 MeV \leq E_{\nu} \leq50 MeV$ (high energy
supernova neutrinos) and high-energy, $50 MeV \leq E_{\nu} \leq
1-2 GeV$ (solar flare-, atmospheric-neutrinos). In our formalism
developed in the present work, we keep the exact momentum
dependence of the operators, since the momentum transfer and the
neutrino-energies involved in the reaction (\ref{H-QPM}) may,
sometimes, be rather high.

Before closing, it is worth mentioning that, the $^{127}\!I$
nucleus constitutes a promising detector also for cold dark matter
candidates \cite{KV97}. This probe relies on the (elastic)
scattering of a cold-dark-matter candidate off the nucleus
$^{127}\!I$ and requires the calculation of the low-lying excited
states of $^{127}\!I$.  Roughly speaking, in the latter case the
required reduced matrix elements are, in general, similar to those
of the reaction $^{127}\!I(\nu,\nu ')^{127}\!I^{\star}$, i.e. the
neutral-current neutrino-scattering. A thorough discussion of this
topic (including calculations for other similar processes) using
the above formalism will be done elsewhere \cite{HaKo02a}.

\section{Summary and Conclusions}

In the present work, we have focused on the theoretical
description of exclusive transitions in neutrino-nucleus
reactions. Assuming that, the initial and final nuclear states
involved have well-defined spin, isospin and parity, and using the
decomposition of the hadronic currents into tensor multipole
operators, we derived closed analytic formulas for the matrix
elements of the principal one-body operators entering the relevant
transition strengths. By employing harmonic oscillator basis,
these matrix elements have been written as products of an
exponential times simple polynomials of even powers in the
momentum transfer q. The coefficients of the polynomials are, in
general, simple and readily calculable numbers as it becomes
evident from Tables 2-4.

The advantage of the above formalism has been illustrated, in the
case of the lowest transition $\frac{5}{2}^+ \ra \frac{3}{2}^+$ of
the currently interesting charged-current reaction
$^{127}\!I(\nu_l,l^-)^{127}\!Xe^{\star}$. The partial rates of
this process are currently evaluated in the framework of the
microscopic quasi-particle phonon model employing for the
single-particle reduced matrix elements our present expressions.
This formalism, in the cases when a phenomenological approach is
adopted, offers the possibility one to make fits for the nuclear
transition strengths in many semi-leptonic processes in nuclei.

\section{Acknowledgements}

This work has been supported in part by the Research Committee of
the University of Ioannina.

\bigskip
\bigskip

\begin{center}
{\bf \Large Appendix A}
\end{center}

The radial integrals of Eq. (\ref{rad-int}) in the case when
harmonic oscillator basis is used can take the following compact
expressions (see Ref. \cite{HaKos}):

(i) For the operator $\theta_L^0(\rho)$ we found \cite{KV97,KRF00}
\beq \Braket{n_1l_1}{j_L(\rho)}{n_2l_2} \, = \, e^{-y}y^{L/2}
\sum_{\mu=0}^{n_{max}}\varepsilon_{\mu}^L \,y^{\mu} , \qquad
y\,=\,(qb/2)^2 \, , \label{rad-int0} \eeq
$$ n_{max}=(N_1+N_2-L)/2,$$
where $N_i=2n_i+l_i$ and the coefficients
$\varepsilon_{\mu}^L(n_1l_1n_2l_2)$ are given by
\beq \varepsilon_{\mu}^L(n_1l_1n_2l_2)\,=\,C \frac{{\pi}^{1/2}}{2}
\sum_{m_1=\phi}^{n_1}\sum_{m_2=\sigma}^{n_2} n!
\Lambda_{m_1}(n_1l_1)\Lambda_{m_2}(n_2l_2)\Lambda_{\mu}(nL),
\label{cf-eps} \eeq
with
$$ n=m_1+m_2+(l_1+l_2-L)/2. $$
In Eq. (\ref{cf-eps}) $C \,=\, b^3 N_{n_1l_1}N_{n_2l_2}/2$ while
the other symbols are explained in Ref. \cite{KV97}.

(ii) The radial matrix elements for the operators $\theta_L^{\pm}$
are given by \cite{HaKos}
\beq
\Braket{n_1l_1}{\theta^{\pm}_L}{n_2l_2} \, = \, e^{-y}y^{(L-1)/2}
\sum_{\mu=0}^{n_{max}}{\zeta}_{\mu}^{\pm}(L) \,y^{\mu},
\label{rad-int-pm}
\eeq
where
\beq {\zeta}_{\mu}^-(L)\,=\, -\frac{1}{2}\left\{\begin{array}{
l@{\quad} l} (n_2+l_2+3/2)^{1/2} \varepsilon_{\mu}^L
(n_1l_1n_2l_2+1)
 + n_2^{1/2}\varepsilon_{\mu}^L (n_1l_1n_2-1l_2+1),&
0\leq \mu< n_{max} \\
(n_2+l_2+3/2)^{1/2}\varepsilon_{n_{max}}^L(n_1l_1n_2l_2+1),  \qquad
\qquad  \mu = n_{max} \\
\end{array} \right.
\label{cf-zetami}
\eeq
\beq {\zeta}_{\mu}^+(L)\,=\, \frac{1}{2} \left\{
\begin{array}{ l@{\quad} l} (n_2+l_2+1/2)^{1/2}
\varepsilon_{\mu}^L(n_1l_1n_2l_2-1)
+  (n_2+1)^{1/2}\varepsilon_{\mu}^L(n_1l_1n_2+1l_2-1),  &
0\leq \mu <n_{max} \\
(n_2+1)^{1/2}\varepsilon_{n_{max}}^L(n_1l_1n_2+1l_2-1),
\qquad\qquad \mu = n_{max}\\
\end{array} \right. \label{cf-etami}
\eeq
and
$$ n_{max}=(N_1+N_2-L+1)/2. $$
The coefficients $\varepsilon_{\mu}^L$ and ${\zeta}_{\mu}^{\pm}$
are calculated using simple codes.

\newpage
 \begin{table}
\begin{center}
 \begin{tabular}{|c|c|c|}  \hline\hline
 & & \\
 {\em Operator} &
 \multicolumn{1}{c|}{\em $\beta$} &
\multicolumn{1}{c|}{\em ${\mathcal P}^J_{\mu} \, ,\quad 0 \leq \mu
\leq{n_{max}}$}
                                             \\ \hline & & \\
 $T_1^J=M^J$ & J & $E_{\mu}^1(J)$  \\ & & \\

 $T_2^J=\Sigma ^J$ & J & $E_{\mu}^2(J)$   \\ & & \\

 $T_3^J={\Sigma}^{\prime J}$ & $J-1$ & $(J+1)^{1/2}E^2_{\mu}(J-1) -
J^{1/2}E_{\mu-1}^2(J+1)$  \\ & & \\

 $T_4^J={\Sigma}^{\prime\prime J}$ & $J-1$ & $J^{1/2}E_{\mu}^2(J-1) +
(J+1)^{1/2}E_{\mu -1}^2(J+1)$  \\ & & \\

 $T_5^J={\Delta}^J$ & $J-1$ & $E_{\mu}^3(L)$  \\ & & \\

 $T_6^J={\Delta}^{\prime J}$ & $J-2$ & $(J+1)^{1/2}E_{\mu}^3(J-1) -
J^{1/2}E_{\mu -1}^3(J+1)$  \\ & & \\

 $T_7^J={\Omega}^J$ & J & $E^4_{\mu}(J)$  \\ & & \\  \hline & & \\
    $\Omega^{\prime J}$ & J -1 & $E^4_{\mu}(J) +
  \frac{1}{2}\Big\{J^{1/2}E_{\mu}^2(J-1) +
   (J+1)^{1/2}E_{\mu -1}^2(J+1)\Big\}$  \\ & & \\ \hline \hline
\end{tabular}
\caption{The coefficients ${\mathcal P}^J_{\mu}$ entering  Eq.
(\ref{gen-pol}) for the seven basic single-body operators entering
the description of any semi-leptonic process in nuclei [see Eqs.
(\ref{opM})-(\ref{opOm}) in the text]. The values for the
parameter $\beta$ of Eqs. (\ref{gen-pol}) and (\ref{nmax}) are
also given. }
\end{center}
\end{table}

 \begin{table}
 \begin{center}
\begin{tabular}{|c|c|c|c|c|c|c|}  \hline\hline
 & & & & & &  \\
 {\em $(n_1l_1)j_1 - (n_2l_2)j_2$} &
 \multicolumn{1}{c|}{\em J} &
 \multicolumn{1}{c|}{\em $\mu = 0$} &
 \multicolumn{1}{c|}{\em ${\mu = 1}$} &
 \multicolumn{1}{c|}{\em ${\mu = 2}$} &
 \multicolumn{1}{c|}{\em ${\mu = 3}$} &
 \multicolumn{1}{c|}{\em ${\mu = 4}$}
\\ \hline & & & & & & \\
$0f_{5/2} - 1p_{3/2}$ & 2 & $\frac{8}{5}\sqrt{\frac{2}{3}}$ &
                       -$\frac{104}{35}\sqrt{\frac{1}{6}}$ &
                             $\frac{8}{35}\sqrt{\frac{2}{3}}$ & &
                                                       \\ & & & & & & \\
                      & 4 & -$\frac{48}{35}$ & $\frac{16}{35}$ & & &
                                                       \\ & & & & & & \\
$1d_{5/2} - 1d_{3/2}$ & 2 & -$\frac{44}{5}\sqrt{\frac{1}{21}}$ &
                          $\frac{376}{35}\sqrt{\frac{1}{21}}$ &
                        -$\frac{152}{35}\sqrt{\frac{1}{21}}$ &
                                     $\frac{16}{35}\sqrt{\frac{1}{21}}$ &
                                                       \\ & & & & & & \\
                      & 4 & $\frac{152}{35}\sqrt{\frac{2}{7}}$ &
                         -$\frac{96}{35}\sqrt{\frac{2}{7}}$ &
                        $\frac{204}{665}\sqrt{\frac{2}{7}}$ & &
                                                       \\ & & & & & & \\
$1f_{5/2} - 1p_{3/2}$ & 2 & -$\frac{4}{5}\sqrt{3}$ &
                           $\frac{40}{35}\sqrt{3}$ &
                       -$\frac{152}{105}\sqrt{\frac{1}{3}}$ &
                      $\frac{16}{105}\sqrt{\frac{1}{3}}$ & \\ & & & & & & \\
                      & 4 & $\frac{136}{105}\sqrt{2}$ &
                        -$\frac{32}{35}\sqrt{2}$ &
                        $\frac{16}{105}\sqrt{2}$ & &
                                                       \\ & & & & & & \\
$1f_{5/2} - 2p_{3/2}$ & 2 & $\frac{16}{5}\sqrt{\frac{3}{7}}$ &
                         -$\frac{40}{7}\sqrt{\frac{3}{7}}$ &
                        $\frac{208}{21}\sqrt{\frac{1}{21}}$ &
                                    -$\frac{232}{105}\sqrt{\frac{1}{21}}$ &
                                    $\frac{16}{105}\sqrt{\frac{1}{21}}$
                                                       \\ & & & & & & \\
                      & 4 & -$\frac{208}{35}\sqrt{\frac{2}{7}}$ &
                         $\frac{592}{105}\sqrt{\frac{2}{7}}$ &
                       -$\frac{176}{105}\sqrt{\frac{2}{7}}$ &
                                    $\frac{16}{105}\sqrt{\frac{2}{7}}$ &
                                               \\ & & & & & & \\ \hline
\end{tabular}
\caption{Coefficients which determine the matrix elements
$\Braket{j_1}{|M ^ J|}{j_2}$ describing the ${\frac{5}{2}}^+\ra
{\frac{3}{2}}^+$ transition of the charge-changing reaction
${}^{127}\!I\, (\nu_l,l^-)\,{}^{127}\!Xe$ in a model space
including the major oscillator shells N=3-5 (see the text).}
 \end{center}
\end{table}
 \begin{table}
 \begin{center}
\begin{tabular}{|c|c|c|c|c|c|c|}  \hline\hline
 & & & & & &  \\
 {\em $(n_1l_1)j_1 - (n_2l_2)j_2$} &
 \multicolumn{1}{c|}{\em J} &
 \multicolumn{1}{c|}{\em $\mu = 0$} &
 \multicolumn{1}{c|}{\em ${\mu = 1}$} &
 \multicolumn{1}{c|}{\em ${\mu = 2}$} &
 \multicolumn{1}{c|}{\em ${\mu = 3}$} &
 \multicolumn{1}{c|}{\em ${\mu = 4}$}
\\ \hline & & & & & & \\
$0f_{5/2} - 1p_{3/2}$ & 2 & -$\frac{8}{3}$ & $\frac{52}{21}$ &
                                               -$\frac{8}{21}$ & &
                                                       \\ & & & & & & \\
                      & 4 & $\frac{24}{7}\sqrt{\frac{1}{5}}$ &
                          -$\frac{8}{7}\sqrt{\frac{1}{5}}$ & & &
                                                       \\ & & & & & & \\
$1d_{5/2} - 1d_{3/2}$ & 2 & -$\frac{22}{3}\sqrt{\frac{2}{7}}$ &
                          $\frac{188}{21}\sqrt{\frac{2}{7}}$ &
                        -$\frac{76}{21}\sqrt{\frac{2}{7}}$ &
                                     $\frac{8}{21}\sqrt{\frac{2}{7}}$ &
                                                       \\ & & & & & & \\
                      & 4 & $\frac{76}{7}\sqrt{\frac{2}{35}}$ &
                         -$\frac{48}{7}\sqrt{\frac{2}{35}}$ &
                        $\frac{8}{7}\sqrt{\frac{2}{35}}$ & &
                                                       \\ & & & & & & \\
$1f_{5/2} - 1p_{3/2}$ & 2 & $2\sqrt{2}$ & -$\frac{20}{7}\sqrt{2}$ &
                        $\frac{76}{63}\sqrt{2}$ & -$\frac{8}{63}\sqrt{2}$ &
                                                       \\ & & & & & & \\
                      & 4 & -$\frac{68}{21}\sqrt{\frac{2}{5}}$ &
                         $\frac{48}{21}\sqrt{\frac{2}{5}}$ &
                        -$\frac{8}{21}\sqrt{\frac{2}{5}}$ & &
                                                       \\ & & & & & & \\
$1f_{5/2} - 2p_{3/2}$ & 2 & -$8\sqrt{\frac{2}{7}}$ &
                          $\frac{100}{7}\sqrt{\frac{2}{7}}$ &
                        -$\frac{520}{63}\sqrt{\frac{2}{7}}$ &
                                     $\frac{116}{63}\sqrt{\frac{2}{7}}$ &
                                    -$\frac{8}{63}\sqrt{\frac{2}{7}}$
                                                       \\ & & & & & & \\
                      & 4 & $\frac{104}{7}\sqrt{\frac{2}{35}}$ &
                         -$\frac{296}{21}\sqrt{\frac{2}{35}}$ &
                        $\frac{88}{21}\sqrt{\frac{2}{35}}$ &
                                    -$\frac{8}{21}\sqrt{\frac{2}{35}}$ &
                                               \\ & & & & & & \\ \hline
\end{tabular}
\caption{Coefficients which determine some reduced matrix elements
$\Braket{j_1}{|\Sigma^J|}{j_2}$. See caption of Table 2 and the
text.  }
\end{center}
\end{table}
 \begin{table}
 \begin{center}
\begin{tabular}{|c|c|c|c|c|c|c|}  \hline\hline
 & & & & & &  \\
 {\em $(n_1l_1)j_1 - (n_2l_2)j_2$} &
 \multicolumn{1}{c|}{\em J} &
 \multicolumn{1}{c|}{\em $\mu = 0$} &
 \multicolumn{1}{c|}{\em ${\mu = 1}$} &
 \multicolumn{1}{c|}{\em ${\mu = 2}$} &
 \multicolumn{1}{c|}{\em ${\mu = 3}$} &
 \multicolumn{1}{c|}{\em ${\mu = 4}$}
\\ \hline & & & & & & \\
$0f_{5/2} - 1p_{3/2}$ & 2 & 0 & $\frac{2}{3}$ &
                                               -$\frac{4}{21}$ & &
                                                       \\ & & & & & & \\
                      & 4 & 0 &
                          -$\frac{4}{7}\sqrt{\frac{1}{5}}$ & & &
                                                       \\ & & & & & & \\
$1f_{5/2} - 1p_{3/2}$ & 2 & $\frac{3}{5}\sqrt{2}$ & -$\frac{22}{35}\sqrt{2}$ &
                      $\frac{122}{315}\sqrt{2}$ & -$\frac{20}{315}\sqrt{2}$ &
                                                       \\ & & & & & & \\
                      & 4 & -$\frac{34}{21}\sqrt{\frac{2}{5}}$ &
                         $\frac{16}{21}\sqrt{\frac{2}{5}}$ &
                        -$\frac{4}{21}\sqrt{\frac{2}{5}}$ & &
                                                       \\ & & & & & & \\
$1f_{5/2} - 2p_{3/2}$ & 2 & 0 & 2$\sqrt{\frac{2}{7}}$ &
                        -$\frac{12}{7}\sqrt{\frac{2}{7}}$ &
                                     $\frac{38}{63}\sqrt{\frac{2}{7}}$ &
                                    -$\frac{4}{63}\sqrt{\frac{2}{7}}$
                                                       \\ & & & & & & \\
                      & 4 & 0 &
                         -$\frac{52}{21}\sqrt{\frac{2}{35}}$ &
                        $\frac{8}{7}\sqrt{\frac{2}{35}}$ &
                                    -$\frac{4}{21}\sqrt{\frac{2}{35}}$ &
                                               \\ & & & & & & \\ \hline
\end{tabular}
\caption{Coefficients which determine some reduced matrix elements
$\Braket{j_1}{|\Delta ^{\prime J}|}{j_2}$. For details see the
text. }
 \end{center}
\end{table}
\newpage
\centerline{\bf \large Figure Caption}

\bigskip

Feynman-diagram of lowest order for: (a) the CC neutrino-nucleus
reactions $\nu_l + {}_{Z}A_{N}\longrightarrow {}_{Z+1}A^*_{N-1} +
l^-$, and (b) the NC neutrino-nucleus processes $\nu + {}_{Z}A_{N}
\longrightarrow {}_{Z}A^*_{N} + \nu^\prime$. The diagrams which
correspond to the anti-neutrino reactions are similar.

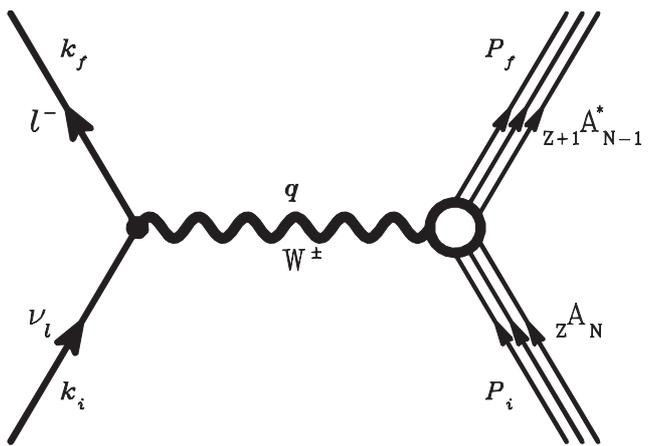
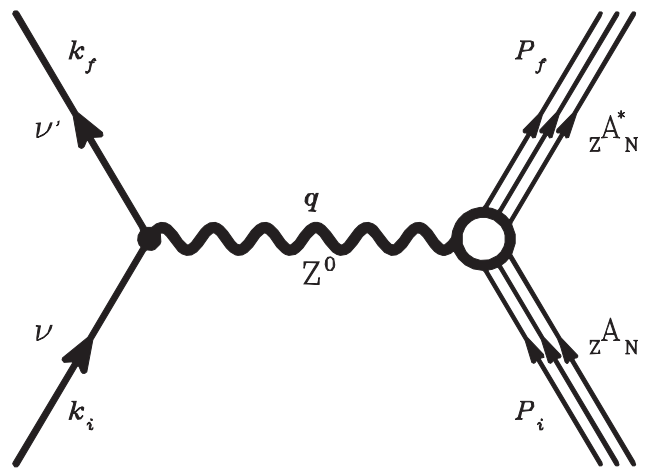
\end{document}